\documentclass[12pt,preprint]{aastex}
\shorttitle{Soft X-ray irradiation of sillicates}
\shortauthors{Ciaravella et al.}

\begin{document}
\title{Soft X-ray Irradiation of Silicates: Implications on Dust Evolution \\ in Protoplanetary Disks}
\author{A. Ciaravella\altaffilmark{1},  C. Cecchi-Pestellini\altaffilmark{1}, Y.-J. Chen\altaffilmark{2}, G.M. Mu\~{n}oz Caro\altaffilmark{3}, \\
C.-H., Huang\altaffilmark{2}, A. Jim\'enez-Escobar\altaffilmark{1} \& A.M. Venezia\altaffilmark{4}}
\altaffiltext{1}{INAF - Osservatorio Astronomico di Palermo, P.za Parlamento 1, 90134 Palermo, Italy; aciaravella@astropa.unipa.it } 
\altaffiltext{2}{Department of Physics, National Central University, Jhongli City, Taoyuan County 32054, Taiwan}
\altaffiltext{3}{Centro de Astrobiolog\'ia (INTA-CSIC), Carretera de Ajalvir, km 4, Torrej\'on de Ardoz, 28850 Madrid, Spain} 
\altaffiltext{4}{ISMN - CNR, Via Ugo La Malfa 153, 90146 Palermo, Italy}

\begin{abstract}
The processing of energetic photons on bare silicate grains was simulated experimentally on silicate films submitted  to soft X-rays of energies up to 1.25 keV. The silicate material  was prepared by means of a microwave assisted sol-gel technique. Its chemical composition reflects the Mg$_2$SiO$_4$ stoichiometry with residual impurities due to  the synthesis method. The experiments were performed using the spherical grating monochromator beamline at the National Synchrotron Radiation Research Center in Taiwan. We found that soft X-ray irradiation induces structural changes that can be interpreted as an amorphization of the processed silicate material. The present results may have relevant implications in the evolution of silicate materials in X-ray irradiated protoplanetary disks.
\end{abstract}


\section{Introduction}
From circumstellar regions of relatively high density where they form, dust grains are ejected into the general interstellar medium. There, they may evolve in response to the interstellar radiation and particle fields and to chemical reactions with interstellar atoms and molecules, changing their chemical  composition and physical structure. Clues about the nature of dust grains are obtained by the numerous astrophysical spectra that invariably identify metal (primarily magnesium  and iron) silicate compounds and carbonaceous matter. Silicate grains, that tend to dominate dust emission in many astrophysical environments, are observed in the cold neutral medium, in comets and protoplanetary disks, and perhaps even in the far distant Universe (e.g., \citealt{D14}). In the diffuse interstellar medium, silicate materials are predominantly amorphous, while crystalline silicates, whose composition is mostly magnesium-rich ~\textendash~ forsterite (e.g., \citealt{J10}) ~\textendash~ represent less than 2 \%  in terms of mass (e.g., \citealt{M07}). Crystalline silicates are frequently observed toward young stars  indicating that in situ formation by thermal annealing or shocks may have occurred \citep{A09}. 

The dust grains that are originally incorporated into protoplanetary disks are essentially of interstellar nature, but have been severely modified by the filter of star formation. Silicate evolution in disks is obviously determined by the changing physical conditions during the evolving disk lifetime. A key question is the extent of such processing that is imprinted in growth, crystallization and settling of dust grains. Surprisingly, as derived from \emph{Spitzer} observations no correlation between the degree of crystallinity in silicate dust and either stellar or disk characteristics have been evidenced (e.g., \citealt{O13}). However,  a few years ago \citet{G09} showed the existence of an anticorrelation between the stellar X-ray activity and the dust crystalline mass fraction, implying dust amorphization in the disk atmospheres. In fact, if the processes affecting the dust within disks have short timescales, and occur constantly other relations between crystallinity and disk/star parameters could be washed out. According to \citet{G09} the inferred amorphization of silicate dust cannot be ascribed straightforwardly to X-rays, but demands an indirect explanation because X-rays carry too little momentum to damage the crystalline structure of the solid. These authors pointed out that stellar wind ions at the disk surface may dominate the whole process.

To validate such a scenario, we consider the effects induced by soft X-ray irradiation on iron-free magnesium silicate materials produced through a sol-gel technique. In the sol-gel process, microparticles or molecules in a solution (sol) agglomerate and under controlled conditions eventually link together to form a coherent network (gel). Generally sol-gel samples are more highly ordered than vapour condensates. We choose to irradiate not pure crystalline solids in order to preserve the interstellar nature of silicates. Other mechanisms acting on the structure of cosmic silicate analogues are thermal annealing \citep{H98}, and exposure to ultraviolet radiation, electrons, and ions  \citep{Y99,C02,De01,De04,B07}. 

We used synchrotron light at the National Synchrotron Radiation Research Center (NSRRC) in Taiwan. Synchrotron sources are ideal because of their high intensity and wide wavelength coverage. In Section \ref{exper} we describe the synthesis of the silicate, its characteristics, and the irradiation experimental setup. The changes in the solid due to the irradiation, and the response to such changes in the infrared spectrum are illustrated in Section \ref{res}. The last Section contains a discussion of the results in the astrophysical context, and our conclusions. 

\section{Synthesis and irradiation of silicates}\label{exper}
The silicate material was prepared by means of a sol-gel method (e.g., \citealt{HW90}).  In order to obtain forsterite silicates (Mg$_2$SiO$_4$), the appropriate amount of the two precursors, hydrated magnesium nitrate $\rm Mg(NO_3)_26H_2O$ and the tetraethyl orthosilicate (TEOS) Si(OC$_2$H$_5$)$_4$ were dissolved in ethanol. The hydrolysis of the TEOS  and the precipitation of the mixed oxide started upon dropwise addition of a 10~M (moles per liter) solution of NaOH. To speed up the condensation process the mixture underwent 50 s of microwave irradiation using a conventional 180~W microwave oven. Filtering and washing with distilled water  was performed to remove 
Na$^+$ and {NO$_3^{\hspace{-0.02cm}-}$} ions. 

\begin{figure}
\centering
\includegraphics[width=8cm]{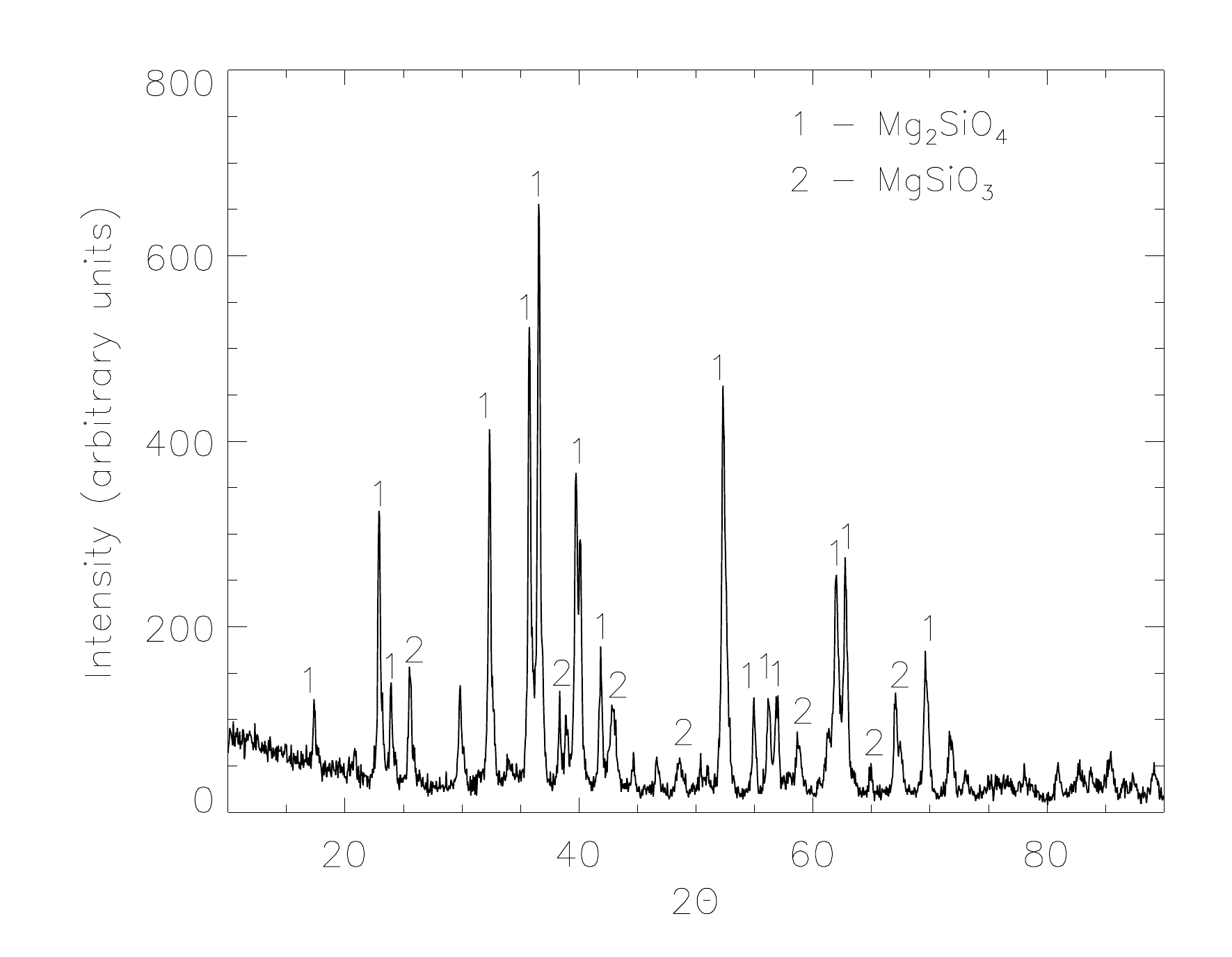}
\caption{X-ray diffraction peak intensities of the silicate after calcination at 800~$^\circ$C.}
\label{s1}
\end{figure}

\begin{figure}
\centering
\includegraphics[width=8cm]{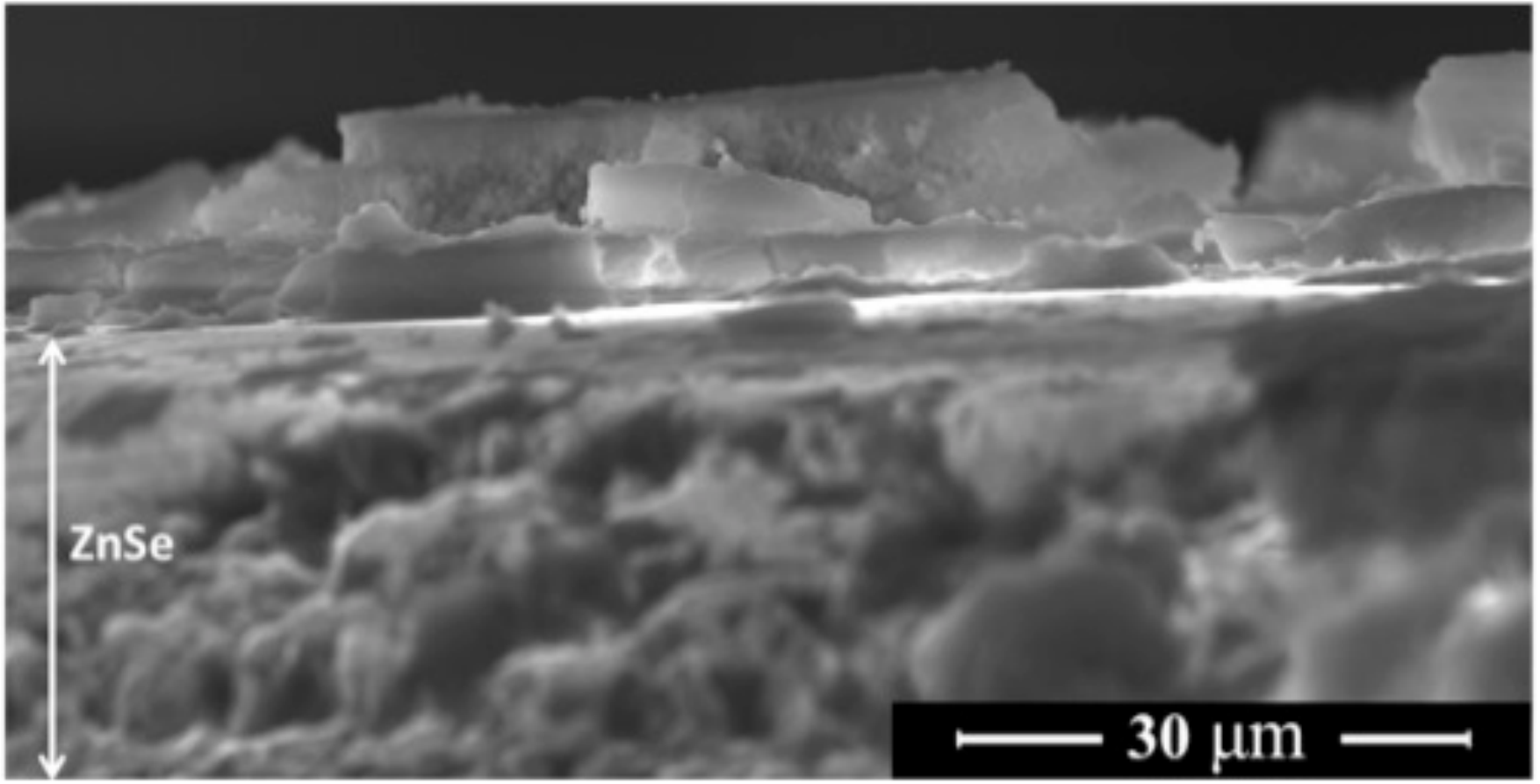}
\caption{SEM image of the synthesized silicate. The region marked by the vertical arrow is the ZnSe window onto which the material has been deposited.}
\label{s2}
\end{figure}

To investigate the structure of the silicate a portion of the obtained powder was calcinated at 800~$^\circ$C and analyzed through X-ray Diffraction (XRD) that showed a pattern  attributable to forsterite (Mg$_2$SiO$_4$) and enstatite (MgSiO$_3$) with a perovskite structure, see Figure~ \ref{s1}. The original silicate suspension was deposited on a ZnSe window of 2.5~cm diameter with a spin-coater devise. The silicate on the window is non homogeneously distributed, as is clear in the Scanning Electron Microscopy (SEM) picture shown in Figure \ref{s2}. The maximum thickness of the silicate film over the ZnSe window is about $\sim 10~\mu$m.

\begin{figure}
\centering
\includegraphics[width=8cm]{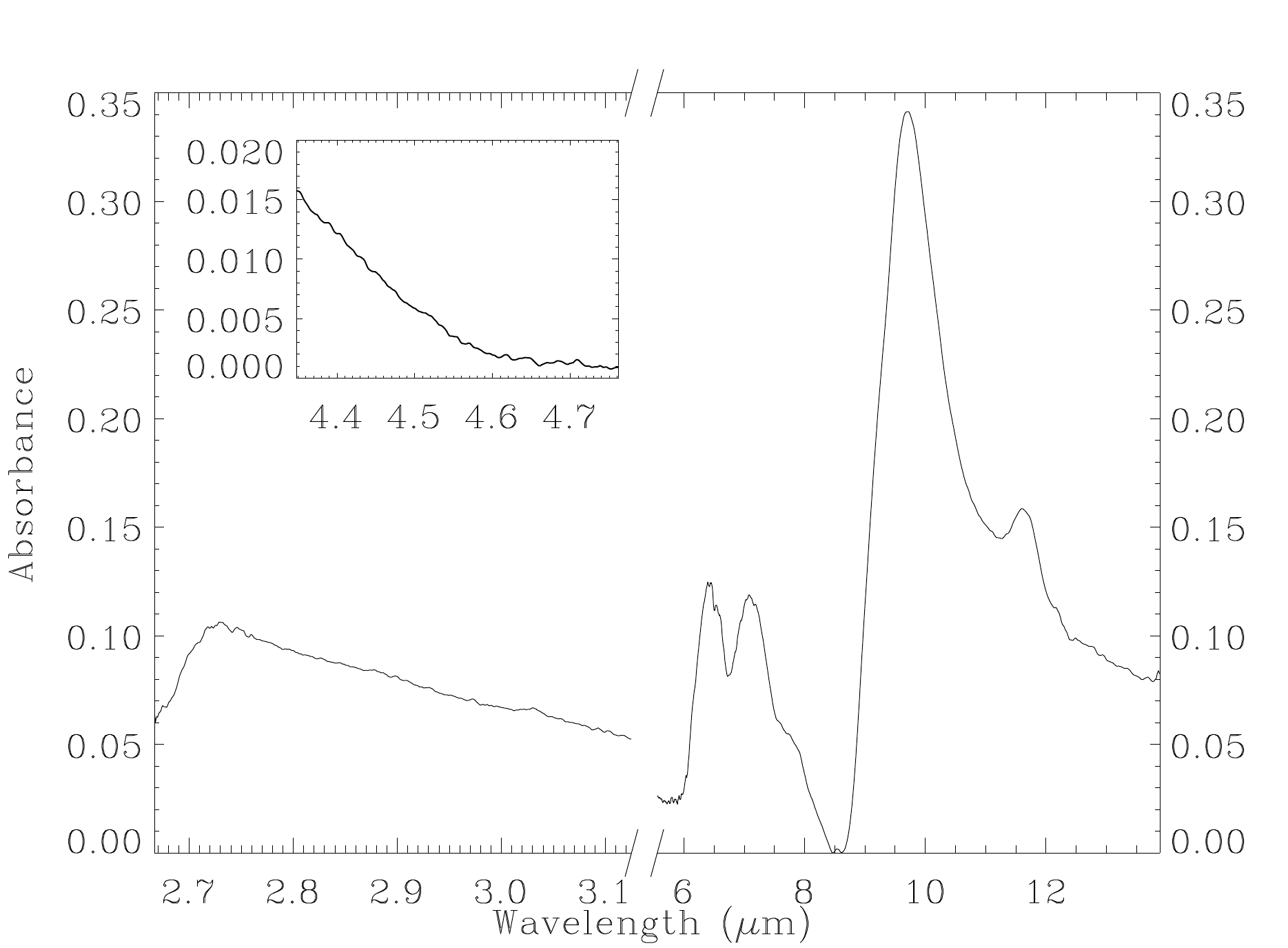}
\caption{Infrared spectra of the silicate at 100 K before the irradiation.}
\label{spec}
\end{figure}

 In Figure~\ref{spec}  the infrared spectra of the  silicate deposited on the ZnSe window shows the frequency of the Si-O stretching mode to be at $9.7~ \mu$m, slightly red-shifted with respect to pure SiO$_2$. As shown by  \citet{J03} the shift increases with the  MgO content. The bands around $7~\mu$m are related to magnesium carbonate \citep{J03} residual from the synthesis procedure, that involves an organic precursor being subsequently dried in air. Its presence was ascertained by the X-ray photoelectron spectroscopy of the material showing a C $1s$ peak at around 290 eV. 
The amount of this contamination was not enough to be detected by XRD analysis. The band at $11.6~\mu$m can be ascribed to the CO$_3^{2-}$ asymmetric bending mode \citep{Ag09} consistent with the presence of the $7~\mu$m band. In silicates produced through gas-phase condensation methods in presence of water traces, the O-Si-H groups  can be responsible for 4.5 and 11.6~$\mu$m bands
(\citealt{S14} and reference therein). The lack of the $4.5~\mu$m band in our sample implies that the $11.6~\mu$m cannot be associated to the  O-Si-H groups. Alternatively, such a band has been assigned to Si$_2$O$_3$ (\citealt{N82} and reference therein). However,  the formation of   Si$_2$O$_3$ requires specific conditions that are not present in the sol-gel method exploited in this work. Synthesis of  Si$_2$O$_3$  may occur through precursors  such as  Si$_2$Cl$_6$ (e.g., \citealt{B91}) or from controlled oxidation of silicon nanowires  \citep{B12}.  The  shoulder at  about $\sim 2.7~\mu$m ($\sim 3670$ cm$^{-1}$) is due to non-associated  Si-OH  contaminants  \citep{M99}.

\begin{figure}
\centering
\includegraphics[width=8cm]{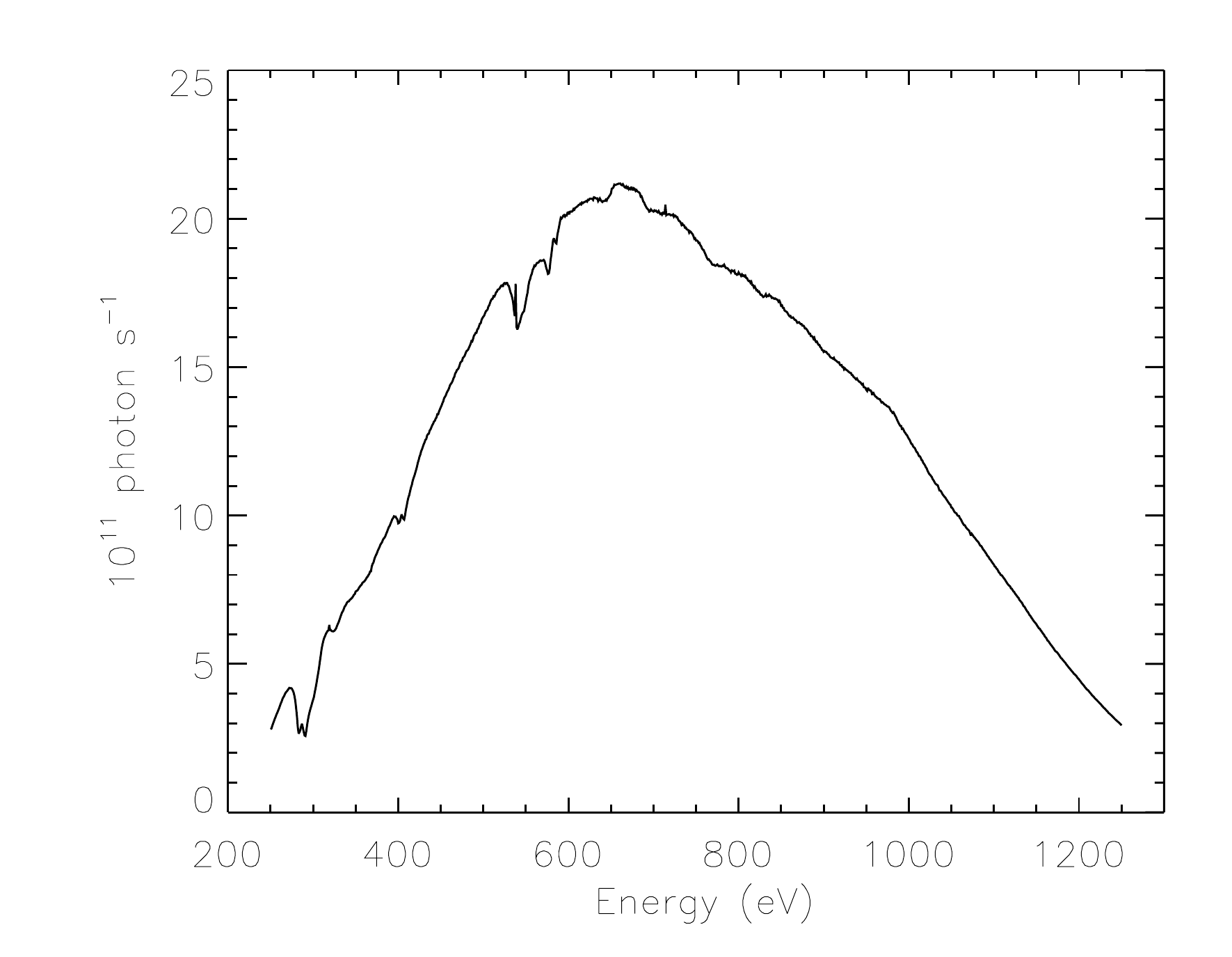}
\caption{X-ray emission spectrum in the range 250 - 1200 eV.  The photon rate in the figure is that for the high flux irradiation.}
\label{xspec}
\end{figure}

The experiments were performed in November 2015 at the NSRRC, using the soft X-ray BL08B beamline providing the broad band (250 - 1250 eV)  emission spectrum reported in Figure~\ref{xspec}. The window coated with silicates was placed in the ultra-high vacuum chamber Interstellar Photo-process System (see \citealt{Ch14} for details), inside the sample holder in contact with a closed-cycle He cryostat (CTI-M350). We performed the experiments twice with different samples of the same synthesized material. The first sample was irradiated using three different photon rates, weak (W,  $\rm 1.67 \times 10^{13} ~ ph ~s^{-1}$), medium (M,  $\rm 6.88 \times 10^{14} ~ ph ~s^{-1}$) and high (H,  $\rm 1.39 \times 10^{15} ~ ph ~s^{-1}$), while the second was processed using only the H rate for the same total energy dose. Table~\ref{tab1} lists the irradiation sequence for the first window. The three columns give the irradiation step, the total irradiation time, in minutes, and the total impinging energy, in eV, at the end of each irradiation step.  The penetration depth of the X-ray photons depends on their energy. If we consider a pure forsterite, photons of 400, 700 and 1200 eV penetrate  $\sim$ 0.27, 0.31 and 1.3~$\mu$m, respectively. Thus, the impinging spectrum is totally absorbed within the first 1.3~$\mu$m.  

A wide X-ray spot was obtained locating the sample about 1~m away from the X-ray focus. A copper mask with a central  hole of the same size as the X-ray spot provided infrared spectra of the irradiated portion only. Since the X-ray  illuminated area is 0.06 cm$^2$, the exploited maximum dose is $\rm \sim 10^{23}~eV~cm^{-2}$.

The samples were placed at 45$^{\circ}$ from the X-ray and infrared beams, cooled to 100~K (to simulate an intermediate position within a protoplanetary disk) and kept at constant temperature during the irradiation. The vacuum inside the chamber during the irradiations was  3 $\times$ 10$^{-10}$ mbar. Infrared spectra of the silicates were taken at the end of each irradiation step. After the irradiation the sample was warmed up  to 300 K  at a rate of  2 K min$^{-1}$.  
\begin{table}
\centering
\caption{Irradiation sequence at 100 K} 
\label{tab1}
\begin{tabular}{c c c}
\hline \hline
Exp.  &  Total Irr. time & Total Energy \\
         &   (minute)        &  ($10^{20}$ eV) \\
\hline 
W1 &  10  &  0.2 \\
W2 &  20  &  0.3 \\
W3 &  30  &  0.4 \\
M1 &  40  &  2.5 \\
M2 &  70  &  9.6 \\
M3 &  100 &  16.2 \\
H1 &  110 &  21.3 \\
H2 &  120 &  26.6 \\
H3 &  150 &  42.2 \\
H4 &  170 &  55.7  \\ \hline
\end{tabular}
\end{table}

\section{Results}\label{res}
\begin{figure}
\centering
\includegraphics[width=8cm]{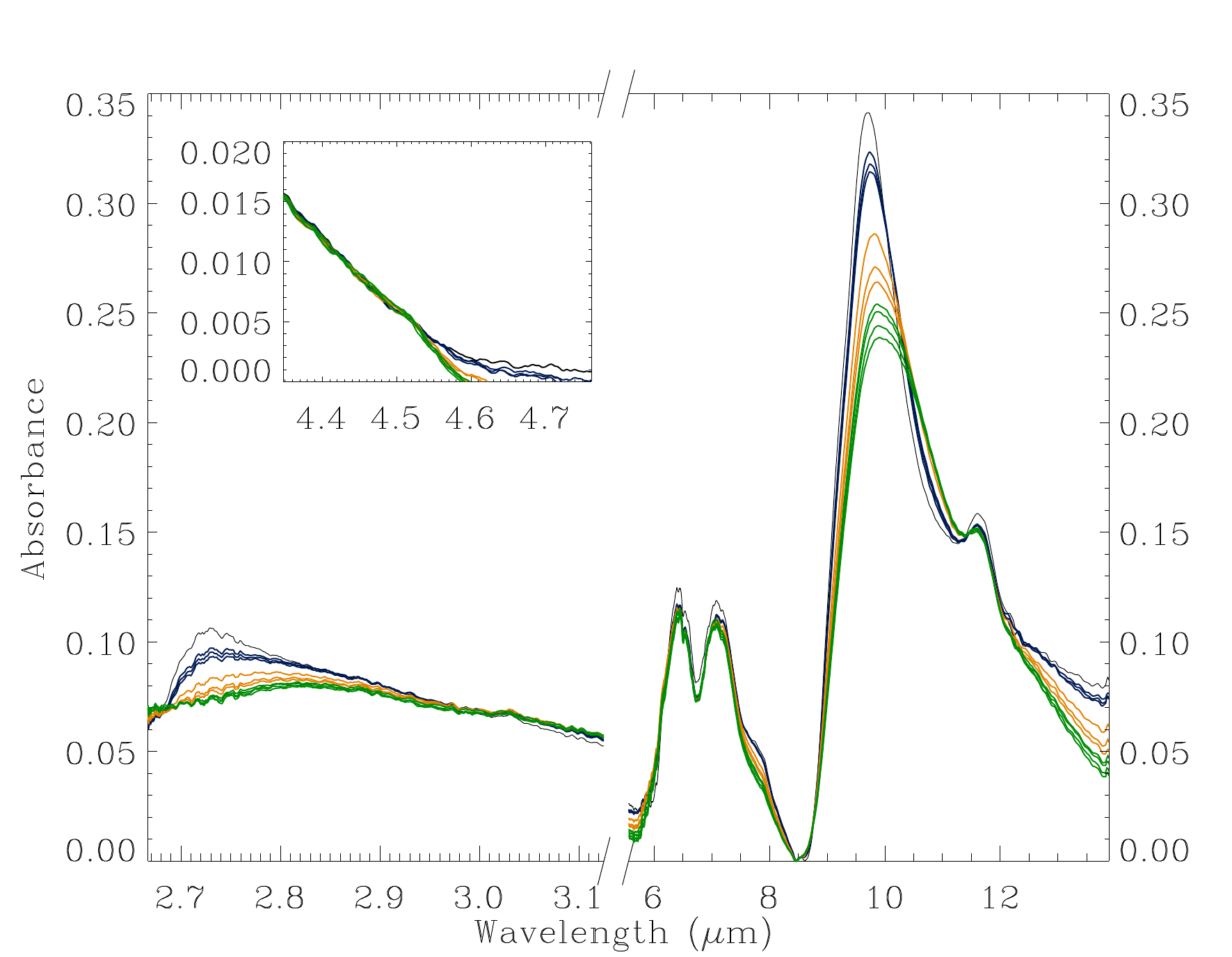}
\caption{Infrared spectra of the silicate. The black line is the silicate spectrum before irradiation. The blue, orange and green lines are the spectra after the irradiations with weak, medium and high X-ray rates.}
\label{s0}
\end{figure}

Figure~\ref{s0} shows the evolution of the infrared spectrum of the material at 100~K during the irradiation experiment. The black curve is the spectrum of the unprocessed sample  (as in Figure~\ref{spec}). 

The spectra in  blue, orange and green are those obtained after the irradiation with weak,  medium  and high fluxes, as listed in Table~\ref{tab1}. The spectra for the second silicate sample show a similar behaviour. X-ray irradiation affects more severely  the bands at 2.7 and 9.7 $\mu$m. A significant decrease of the first is most probably due to the loss of the OH group, as suggested by the photo-desorption of the masses 1 and 16 
detected by quadrupole mass spectrometer during the irradiation.  The Si-O stretching mode profile becomes weaker and broader. The band peak position also shows a redshift displacement of about $0.2~\mu$m. The effects of the X-rays on the silicate band increase with the deposited energy. The largest jump  between the bottom blue and the top yellow profiles corresponds to the largest difference  (about a factor 6)  in the deposited energy, see Table~\ref{tab1}. 
Figure~\ref{s0} shows a saturation of the profile modifications with the deposited X-ray energy.
The modifications of the line profile caused by the X-rays are permanent. 

\section{Analysis of the Silicate Evolution}\label{AN}

\begin{figure}
\centering
\includegraphics[width=7cm]{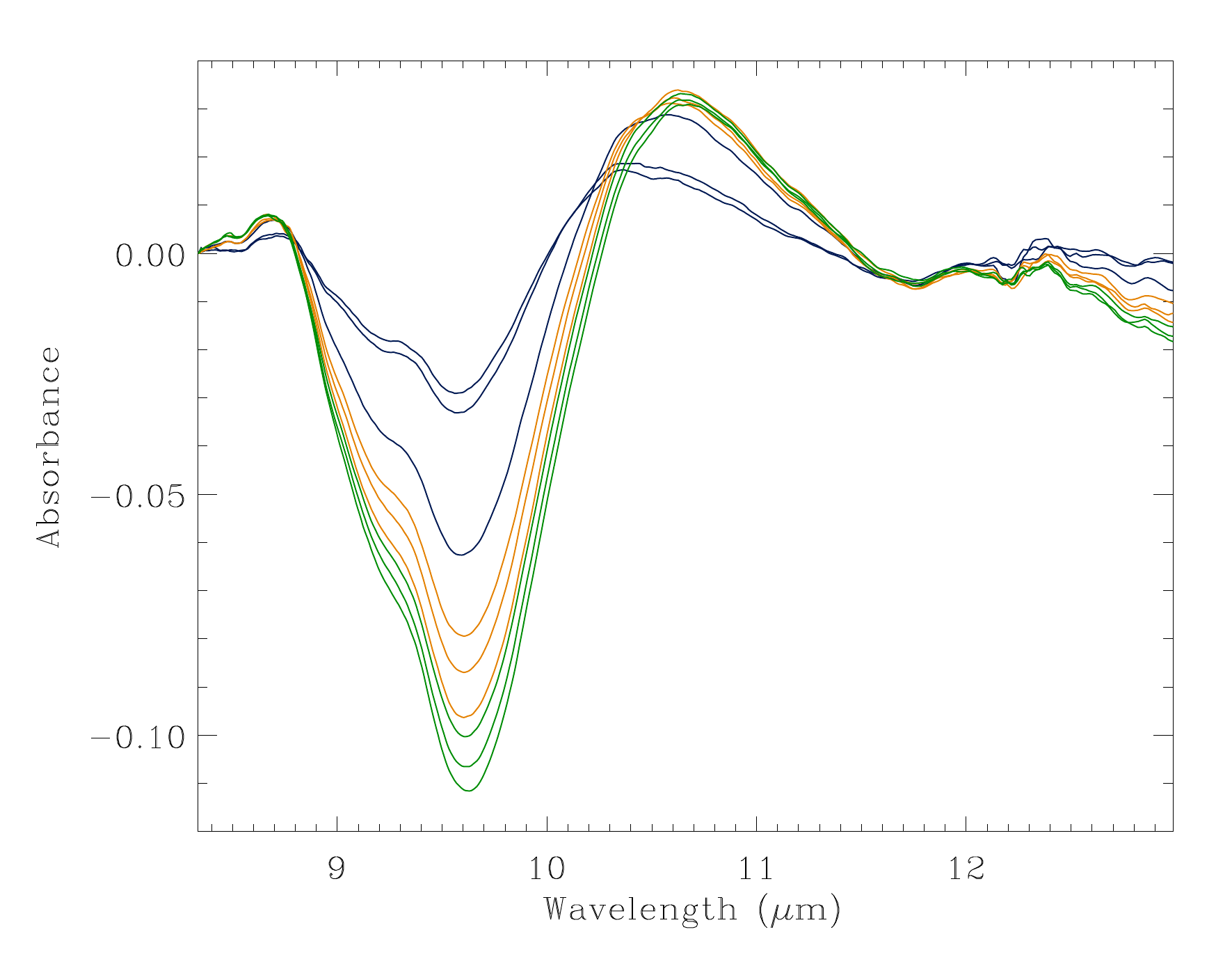}
\caption{Infrared difference spectra obtained subtracting from the spectra after each irradiation step the spectrum of the silicate sample. The  blue, orange and green lines are the difference spectra for the irradiations with weak, medium and high X-ray rates.}
\label{dif}
\end{figure}

The difference between the infrared spectrum after each irradiation step and the sample spectrum for the Si-O stretching band is
shown in Figure~\ref{dif}. 
Blue, orange and green, as in Figure~\ref{s0},  correspond to weak, medium and high X-ray rate irradiations. 
While the peak at $9.7~\mu$m decreases its shoulder around $10.7~\mu$m increases with the irradiation. The deep at $11.6~\mu$m  
indicates a small decrease in the carbonate feature as well. 

\begin{figure}
\centering
\includegraphics[width=8cm]{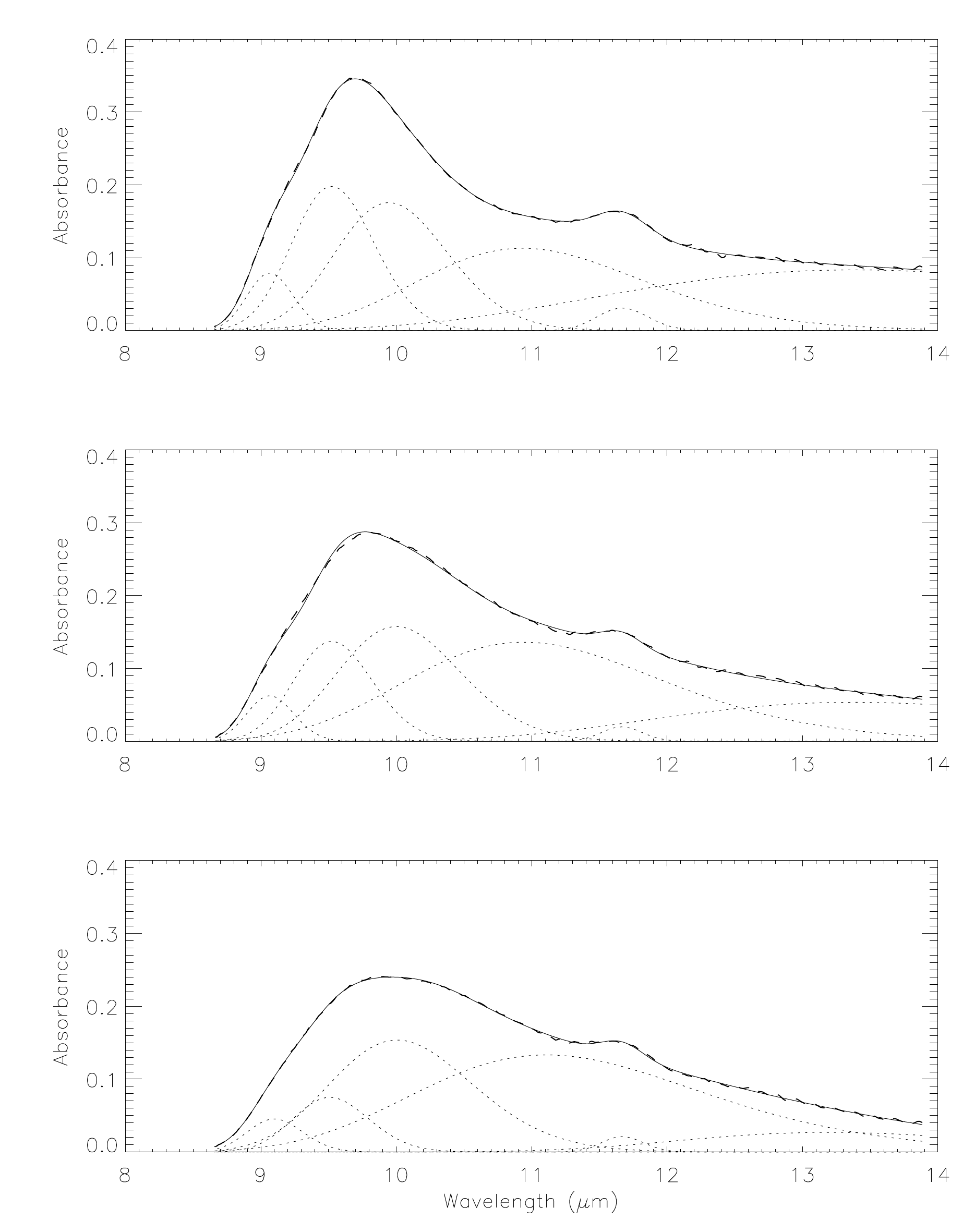}
\caption{Deconvolution of the Si-O stretching band at $9.7~\mu$m with six Gaussian curves before (top panel), during (middle panel, M1 in Table~\ref{tab1}), and after the irradiation (bottom panel). The peaks of the six Gaussian curves in the profile of the silicate sample (top panel)  are at  9.1, 9.5, 9.9, 10.9, 11.6 and 13.4~$\mu$m. }
\label{gau}
\end{figure}

In order to understand the behaviour of the feature at $9.7~\mu$m  the band has been deconvolved using a multiple Gaussian fit.  From the shape of the feature,  six components corresponding to peaks and shoulders in the 
band profile have been identified. The best fit of the $9.7~\mu$m  band was obtained with six components. In the fitting procedure the peaks of the Gaussian curves have been fixed  within $\pm$ 0.1~$\mu$m leaving free the remaining parameters. The deconvolution of the Si-O band profile of the silicate sample before ($\chi^2 = 1.7$), during (M1 in Table~\ref{tab1}) ($\chi^2 = 5.2$) and after  ($\chi^2 = 2.2$) the irradiation are shown in Figure~\ref{gau}. 

The Gaussian curves in this figure indicate that  the components at $\sim$ 9 and 9.5$~\mu$m decrease with the irradiation time, while those at  $\sim 9.9$ and $10.9~\mu$m increase becoming about 25\% wider at the end of the irradiation. Both are shifted toward larger wavelengths with the $10.9~\mu$m component being red-shifted of about 0.2$~\mu$m. The intensity of the component at  $\sim$ 11.6$~\mu$m is reduced.  Finally, the integrated absorbance of the Si-O band as obtained from the six Gaussian components decreases by about 20\% with the irradiation. 

\section{Discussion and Astrophysical Implications}\label{DC}

The first evident conclusion of this work is that soft X-ray irradiation does modify the structure of the silicate sample. The Si-O stretching  band shifts by  about $0.2~\mu$m toward longer wavelengths, becoming weaker and broader along with the irradiation time. Such a change in the profile shows some similarities with the first evolutionary stages in the bands of amorphous magnesium silicate smokes subjected to thermal annealing, as shown by
\citet{H98}. The deconvolution of the Si-O band profile in six Gaussian components (discussed in the previous Section, see Figure~\ref{gau}) shows an increase and broadening of some components and a sharp decrease in others. Such a trend differs from that observed by \citet{H98}, in which there is a steady rise of the components. We thus interpret the variations as a loss of the residual order of the silicate sample rather than a local ordering due to thermal annealing, induced by X-rays at the microscopic level. 

Moreover, the red-shift in the silicate profile observed during  the annealing by \citet{H98} is also detected  in the amorphization of silicates induced by high pressure conditions (e.g., \citealt{W90}). The Si-O bond being the most covalent bond in silicates is the less resistant link in the network structure to amorphization (see the discussion in \citealt{T05}).
 
Amorphization is known to be induced by direct collisions of incident low energy (keV) ions with the atoms forming the silicate lattice. This produces a displacement cascade at the end of each ion track, causing the local collapse of the crystalline structure (e.g., \citealt{De01,De04}). Using ions of much higher energies,  triply ionized 10 MeV Xe ions,  \citet{B07} irradiated samples of forsterite single crystals. At this energy, a large fraction of the total stopping power is due to the inelastic collisions between bound electrons in the medium and the ion moving through it (swift heavy ion), and decreases with penetration depth as the ion slows down. TEM studies revealed that in the \citet{B07} experiment the sample was amorphized up to a depth of about  half of the ion range. During the impact, a cylindrical pressure wave expands from the ion track ejecting molecules from the surface of the solid and creating a nanometre-sized track of dense electronic excitations that can be in part converted into atomic motion. The cylinder along the ion track is the region of amorphization of the material. Recently, \citet{G16} detected the amorphization of polycrystalline silicates embedded in an organic matrix after intense hard (29.4 keV) X-ray irradiation. Matrix-free crystalline silicate grains did not exhibit amorphization, while none of the amorphous samples in any configuration underwent crystallization. \cite{G16} attributed the amorphization mechanism to the electrostatic discharges leading to the breakdown of the dielectric lattice, such as in Coulomb explosions. In the process the role of the organic matrix is to allow charge build up by secondary electrons. 

In the present experiment we exploit soft X-rays  of energies ($250 - 1250$ eV) compatible with stellar emission. The  mean-free paths of photo-electrons are then significantly reduced compared to the ones characteristic of  photo-electrons produced by X-rays  $30 - 100$ times more energetic as those exploited by \citet{G16}. Thus, the electron energy deposition is local. The cascade accumulation of the outer valence electrons responsible for chemical bonding, that are easily stripped from atoms, provides coupling of the electronic excitation energy into atomic motion. 

The interaction of the photo-electrons is primarily plasmon-excitation that leads to a decay into electron-hole pairs. The number of secondary electrons produced through such a mechanism depends on  the photo-electron energy and the mean energy for electron-hole pair  production (for a given material). The  pair-creation energy is generally a few tens of eV (e.g.,  \citealt{V96}), which implies at least 10 secondary electrons per primary photo-electron. We estimate the fractional ionization rate in the sample (see  \citealt{Cia16}) computing the number of photons absorbed inside a volume defined by the X-ray spot and the maximum depth of $\sim 1~\mu \rm m$. The rate of photons absorbed inside such a volume is about $1.3 \times 10^{13}$,  $5.4 \times 10^{14} $,  and $\rm 1.1 \times 10^{15} ~ ph ~s^{-1}$ during the weak, medium and high flux irradiation, respectively.  Assuming 
an upper limit to the density as for the pure Mg$_2$SiO$_4$ of 3.27 g~cm$^{-3}$, the volume contains about $10^{17}$ molecules. Including the secondary electron cascade, the resulting fractional ionization rates (s$^{-1}$) are 0.001,  0.05, and 0.1 for the weak, medium and high photon rates, respectively. Thus, the observed effect is not related to the occurrence of overlarge local electron densities. We explored two orders of magnitude in the X-ray photon rate (and correspondingly in the fractional ionization rate) and we did detect changes in the band profile even for the lowest irradiation step (3.5 $\times$ 10$^{20}$ eV cm$^{-2}$). 

The results of our experiment may be relevant to silicate evolution in protoplanetary disks surrounding typical T Tauri stars. A recent model of a disk embedding  a star with mass of  0.5 M$_{\sun}$ radius of 2 R$_{\sun}$, effective temperature of  4000 K and X-ray luminosity of 10$^{30}$ erg s$^{-1}$ shows that X-rays dominate in regions where emission at lower energies are inhibited \citep{W12}. In regions of the  disk at a temperature of 100 K, the X-ray flux is   $\sim 6.2 \times 10^{9}$ eV cm$^{-2}$ s$^{-1}$ at 1 AU and Z/R = 0.08,  and $\sim 6.2 \times 10^{13}$ eV cm$^{-2}$ s$^{-1}$ at 10 AU and Z/R = 0.26.  At these locations inside the disk the same energies experienced by our sample after  the first ($3.3 \times 10^{20}$ eV cm$^{-2}$) and the last irradiation step ($\rm \sim 9.3 \times 10^{22}$ eV cm$^{-2}$) would require  times of $(0.02 - 5) \times 10^5$~yr  at 1 AU and  $0.2 - 50$~yr, at 10 AU, respectively. The expected irradiation times can be even shorter considering that young stellar objects may have much higher X-ray luminosities, and intense flare activity during which the X-ray emission can be significantly higher with respect to the quiescent emission considered in the \citet{W12} model (see e.g., \citealt{F05}). The dust processing timescale by X-rays can be thus short enough to explain the lack of correlation with any other physical parameters.

The results of this work show that soft X-rays alter the silicate structure. Our interpretation of such modification has been ascribed to a  loss of local residual order (an amorphization) of the silicate.  Such an interpretation will be further tested by  future experiments on pure crystalline samples.
 
\acknowledgments
We thank the referee, Dr J.A. Nuth, for the useful and constructive comments to our work. We are grateful to Prof. G. Marc\`{i} (DEIM, University of Palermo) for having performed the SEM of the sample. We acknowledge support from INAF through the Progetto Premiale: "A Way to Other Worlds" of the Italian Ministry of Education, University, and Research. This work was also financially supported by the Autonomous Region of Sardinia, Project CRP 26666 (Regional Law 7/2007, Call 2010), the MOST grants MOST103-2112-M-008-025-MY3 (Y.-J.C.), Taiwan, and project AYA-2011-29375,AYA-2014-60585-P of Spanish MINECO.


\begin{thebibliography}{} 
\bibitem[\'{A}brah\'{a}m et al.(2009)]{A09} \'{A}brah\'{a}m,  P., Juh\'{a}sz, A., Dullemond, C.P. et al. 2009, Nature, 459, 224
\bibitem[Aguiar et al.(2009)]{Ag09} Aguiar, H.,  Serra, J., Gonz‡lez, P. \& Le\'on, B. 2009, JNCS,  355, 475
\bibitem[Bashauti et al.(2012)]{B12} Bashauti, M.Y., Sardashi, K., Ristein, J.,  Christiansen, S. H. 2012,  PCCP, 14, 11877
\bibitem[Belot et al.(1991)]{B91}Belot, V., Corriu, R.J.P., Leclercq, D., Lef\`evre, P., Mutin, P.H., and Vioux, A. 1991, JNCS, 127, 207 
\bibitem[Bringa et al.(2007)]{B07} Bringa, E.M., Kucheyev, S.O., Loeffler, M.J. et al. 2007, \apj, 662, 372
\bibitem[Carrez et al.(2002)]{C02} Carrez, P., Demyk, K., Leroux, H., Cordier, P., Jones, A. P., D'Hendecourt, L. 2002, M\&PS, 37, 1615
\bibitem[Chen et al.(2014)]{Ch14} Chen, Y.-J. , Chuang, K.-J. , Mu\~{n}oz Caro, G.M., Nuevo, M., Chu, C.-C. , Yih T.-S. , Ip, W.-H. \& Wu C.-Y. R., 2014, \apj, 781, 15
\bibitem[Ciaravella et al.(2016)]{Cia16} Ciaravella, A., Chen, Y.-J, Cecchi-Pestellini, C., Jim\'{e}nez Escobar, A.,  Mu\~{n}oz Caro, G.M., Chuang, K.-J. \& Huang, C.-H.   2016, \apj, 819, 38
\bibitem[Demyk et al.(2001)]{De01} Demyk, K., Carrez, Ph., Leroux, H., Cordier, P., Jones, A.P., Borg, J., Quirico, E., Raynal, P. I. \& d'Hendecourt, L. 2001, \aap, 368, L38
\bibitem[Demyk et al.(2004)]{De04} Demyk, K., d'Hendecourt, L., Leroux, H., Jones, A.P. \& Borg, J. 2004, \aap, 420, 233
\bibitem[Dwek et al.(2014)]{D14} Dwek, E., Staghun, J., Arendt, R.G., Kovacs, A., Su, T. \& Benford, D.J., 2014, \apj, 788, L30
\bibitem[Favata et al.(2005)]{F05} Favata, F., Flaccomio, E., Reale, F., Micela, G., Sciortino, S., Shang, H., Stassun, K.G. \& Feigelson, E.D. 2005, \apjs, 160, 469
\bibitem[Gavilan et al.(2016)]{G16} Gavilan, L., J\"{a}ger, C., Simionovici, A. et al. 2016, \aap, 587, A144
\bibitem[Glauser et al.(2009)]{G09} Glauser, A.M., G\"{u}del, M., Watson, D.M., et al. 2009, \aap, 508, 247
\bibitem[Hench \& West(1990)]{HW90} Hench, L.L. \& West, J.K. 1990, Chem. Rev., 90, 33
\bibitem[Hallenbeck et al.(1998)]{H98} Hallenbeck, S.A., and Nuth III, J. A. , and Daukantas, P.L. 1998, Icarus, 131, 198
\bibitem[J\"{a}ger et al.(2003)]{J03} J\"{a}ger, C., Dorschner, J., Mutschke, H., Posch, Th. \& Henning, Th. 2003, \aap, 408, 193
\bibitem[Juh\'{a}sz et al.(2010)]{J10} Juh\'{a}sz, A., Bouwman, J., Henning, Th., et al. 2010, \apj, 721, 431
\bibitem[Min et al.(2007)]{M07} Min, M., Waters, L.B.F.M., de Koter, A., Hovenier, J.W., Keller, L.P. \& Markwick-Kemper, F. 2007, \aap, 462, 667
\bibitem[Morimoto et al.(1999)]{M99}Morimoto, Y., Nozawa, S.,\& Hosono, H. 1999, Phys. Rev. B, 59, 4066
\bibitem[Nuth \& Donn(1982)]{N82}Nuth, J. A.,  Donn, B. 1982, \apj, 257, L103
\bibitem[Oliveira et al.(2013)]{O13} Oliveira, I., Mer\'{i}n, B., Pontoppidan, K.M. \& van Dishoeck, E.F. 2013, \apj, 762, 128 
\bibitem[Sabri et al.(2014)]{S14}Sabri, T., Gavilan, L., J\"{a}ger, C., Lemaire, J. L.;,Vidali, G., Mutschke, H., Henning, T. 2014, \apj, 780, 180
\bibitem[Trachenko et al.(2005)]{T05} Trachenko, K., Pruneda, J.M., Artacho, E. \& Dove, M.T. 2005, Phys. Rev. B,  71, 184104
\bibitem[Vidal et al.(1996)]{V96}Vidal, R., Baragiola, R.A. \& Ferr\'on, J. 1996, J. Appl. Phys., 80, 5653
\bibitem[Walsh et al.(2012)]{W12} Walsh, C., Nomura, H.,  Millar, T.J. \& Aikawa, Y. 2012, \apj, 747, 114
\bibitem[Williams et al.(1990)]{W90} Williams, Q., Knittle, E., Reichlin, L., Martin, S. \& Jeanloz, R. 1990, \jgr, 95, 21549
\bibitem[Yen et al.(1999)]{Y99} Yen, A. S.,  Grunthaner, F. J.;, Kim, S. S.,  Hecht, M. H. 1999,  LPSC, 30, 1924
\end{thebibliography}
\end{document}